\documentclass[sigconf]{acmart}

\AtBeginDocument{%
  }

\setcopyright{acmlicensed}
\copyrightyear{2026}
\acmYear{2026}
\acmDOI{XXXXXXX.XXXXXXX}
\acmConference[EASE 2026]{The 30th International Conference on Evaluation and Assessment in Software Engineering}{9–12 June, 2026}{Glasgow, Scotland, United Kingdom}

\acmISBN{978-1-4503-XXXX-X/2018/06}

\usepackage{multirow}  

\newcommand{\AUC}{\textit{AUC}}
\newcommand{\AUCB}{\AUC~}

\newcommand{\nn}{\textit{n}}
\newcommand{\nnB}{\nn~}

\newcommand{\FPR}{\textit{FPR}}
\newcommand{\FPRB}{\FPR~}
\newcommand{\PPV}{\textit{PPV}}

\newcommand{\TPR}{\textit{TPR}}
\newcommand{\TPRB}{\TPR~}
\newcommand{\TNR}{\textit{TNR}}

\newcommand{\TP}{\textit{TP}}
\newcommand{\FP}{\textit{FP}}
\newcommand{\TN}{\textit{TN}}
\newcommand{\FN}{\textit{FN}}
\newcommand{\AP}{\textit{AP}}
\newcommand{\AN}{\textit{AN}}

\newcommand{\TNB}{\textit{\TN~}}
\newcommand{\FNB}{\textit{\FN~}}
\newcommand{\APB}{\textit{\AP~}}
\newcommand{\ANB}{\textit{\AN~}}

\newcommand{\THETITLE}{Evaluating Software Defect Prediction Models via the Area Under the ROC Curve Can Be Misleading}

\begin{document}

\title[\THETITLE]{\THETITLE}
\author{Luigi Lavazza}
\email{luigi.lavazza@uninsubria.it}
\orcid{0000-0002-5226-4337}
\affiliation{%
	\institution{Universit\`a degli Studi dell'Insubria}
	\city{Varese}
	\country{Italy}
}
\author{Gabriele Rotoloni}
\email{grotoloni@uninsubria.it}
\orcid{0000-0003-2046-0090}
\affiliation{%
	\institution{Universit\`a degli Studi dell'Insubria}
	\city{Varese}
	\country{Italy}
}
\author{Sandro Morasca}
\email{sandro.morasca@uninsubria.it}
\orcid{0000-0003-4598-7024}
\affiliation{%
	\institution{Universit\`a degli Studi dell'Insubria}
	\city{Varese}
	\country{Italy}
}

\renewcommand{\shortauthors}{Lavazza et al.}

\begin{abstract}
\emph{Background}:
Receiver Operating Characteristic (ROC) curves are widely used to evaluate the performance of Software Defect Prediction (SDP) models that estimate module fault-proneness, i.e., the probability that a module is faulty.
A ROC curve maps a model's performance in terms of True Positive Rate and False Positive Rate for any possible threshold set on fault-proneness.
The Area Under the ROC Curve (\AUC) summarizes the performance of a model across all possible thresholds.
Traditionally, ROC curves completely above the bisector of the ROC space are considered better than random, and high \AUCB values are associated with good performance.
\\
\emph{Aim}:
We investigate whether these beliefs are correct, hence if SDP model evaluation based on ROC curves and \AUCB is reliable.
\\
\emph{Method}:
We decorate ROC curves by highlighting the points corresponding to threshold values.
We also represent True Positive Rate and False Positive Rate as functions of the threshold.
Thus, we can evaluate whether a model classifies both faulty and non-faulty modules better than the random model.
\\
\emph{Results}:
We show that commonly used evaluation criteria may lead to wrong conclusions.
\\
\emph{Conclusions}:
A high value of \AUCB does not guarantee that both the True Positive Rate and the False Positive Rate of a model are better than the random model's for all possible thresholds.
Either decorated ROC curves or alternative representations are needed to appreciate all the relevant aspects of SDP models.
\end{abstract}




\maketitle

\section{Introduction}\label{sec:into}

Module fault-proneness, i.e., the probability that a software module is faulty, can be estimated via 
Software Defect Prediction (SDP) models. To classify a module as faulty or not faulty, a threshold $t\!\in\![0,1]$ is set on fault-proneness: a module is classified as faulty (positive) if its fault-proneness is above $t$ and non-faulty (negative) otherwise.
A Receiver Operating Characteristic (ROC) curve plots the values of \TPRB (the True Positive Rate) against \FPRB (the False Positive Rate) obtained from a dataset by using all possible threshold values $t$ on a fault-proneness model~\cite{FawcettPRL2006}.

The Area Under the ROC Curve (\AUC) is widely used to evaluate SDP models~\cite{moussa2022use}.
Evaluations rely on the following well-established assumptions: i) a ROC curve that is entirely above the bisector of the ROC space performs better than the random classifier, since  the bisector is the ROC curve of the random classifier; ii) high values of \AUCB are associated with good performance; hence, given two models, the one with the higher \AUCB is considered preferable.

Although ROC curves are used to represent the performance obtained by varying the cut-off threshold, ROC curves do not use or represent the value of the threshold itself. Once the modules have been ranked by decreasing fault-proneness, it is possible to draw the ROC curve or to compute  \AUC~\cite{ferri2005modifying} based only on module ranking and the actual faultiness of modules.

One could wonder if ignoring thresholds is a correct practice.
In this paper, we reconsider ROC curves by accounting for the threshold value that generated each one of the curve points.
In this way, it becomes apparent that many models, whose performance is considered good when evaluated via traditional criteria, actually do not perform really well. Specifically, it is often the case that either \TPRB or \FPRB is good, i.e., many models are either good at classifying defective modules or not defective ones, but not both.

The contribution of this paper is of a methodological nature. The problems with ROC curves and \AUCB derive from the very nature of ROC curves.
An empirical study shows that misleading indications are produced frequently.
To overcome these problems, we suggest some methodological guidelines.

The paper is organized as follows.
Section~\ref{sec:background} provides some background concerning ROC curves, \AUCB and the random classifier.
Section~\ref{sec:foundations} presents the conceptual observations that explain why the traditional interpretation of ROC curves and  \AUCB can be misleading.
Section~\ref{sec:empirical} shows examples of real-life SDP models that are considered good according to traditional criteria, but actually do not perform well.
Section~\ref{sec:imbalance} discusses the effect of imbalance on ROC curves.
Section~\ref{sec:study} illustrates an empirical study that evaluates how frequently traditional criteria can be misleading.
Section~\ref{sec:usage} suggests how to evaluate SDP models using ``threshold-aware'' ROC curves.
Section~\ref{sec:related} discusses the related work.
Finally, Section~\ref{sec:conclusions} draws some conclusions.

\section{Background}\label{sec:background}
In this section, we summarize the basics of performance metrics, ROC curves, and \AUC, and the related assumptions.

\subsection{The Confusion Matrix}

The performance of a binary classifier on a set of \nnB instances is usually assessed based on a $2\!\times\!2$ ``confusion matrix'' that shows how many of those \nnB items are correctly and incorrectly classified. For consistency with the literature, we also say that a faulty module is positive and a non-faulty one is negative.
As Table~\ref{tab:CM} shows, the cells of a confusion matrix contain the numbers of modules that are: correctly estimated negative (True Negatives \TN); incorrectly estimated negative (False Negatives \FN); incorrectly estimated positive (False Positives \FP); and correctly estimated positive (True Positives \TP).
The numbers of actually negative \ANB and positive modules \APB depend exclusively on the dataset and not on the specific binary classifier.
An especially important characteristic of a dataset is the prevalence of the positive class, i.e., $\rho=\AP/n$.

\begin{table}[hbt]
\centering
\caption{Confusion matrix}
\label{tab:CM}
\begin{tabular}{c c|c|c|} 
& \multicolumn{1}{c} {} & \multicolumn{2}{c}{Actual}  \\
& \multicolumn{1}{c} {} & \multicolumn{1}{c} {Negative} & \multicolumn{1}{c} {Positive} \\
\cline{3-4}
\multirow{2}{*} {\rotatebox[origin=c]{90}{Estimated}} & Negative & \TN & \FN \\
&  & (True Negatives) & (False Negatives) \\
\cline{3-4}
& Positive & \FP & \TP \\ 
&  & (False Positives) & (True Positives) \\
\cline{3-4}
& \multicolumn{1}{c} {} & \multicolumn{1}{c} {\ANB = \TNB + \FP} & \multicolumn{1}{c} {\APB = \FNB + \TP} \\
& \multicolumn{1}{c} {} & \multicolumn{1}{c} {(Actual Negatives)} & \multicolumn{1}{c} {(Actual Positives)} \\
\end{tabular}
\end{table}

Based on a confusion matrix, several performance metric can be computed: \TPR=\TP/\APB(the True Positive Rate, alias Recall), \FPR= \FP/\ANB (the False Positive Rate), \PPV=\TP/(\TP+\FP) (the Positive Predicted Value, alias Precision), etc.

\subsection{The Receiver Operating Characteristic (ROC) Curve}\label{subsec:roc}

In this paper, we consider SDP models that estimate module fault-proneness, i.e., the probability that a software module is faulty (i.e., probabilistic scoring classifiers~\cite{FawcettPRL2006}).

To classify a module as faulty or not, a threshold $t \in [0,1]$ is set on fault-proneness: a module is classified as faulty (positive) if its fault-proneness is above $t$ and non-faulty (negative) otherwise.

Many different binary classifiers can be obtained by varying $t$, with different classification performance. Each of them is characterized by a different value for the pair $\langle\FPR,\TPR\rangle$.
A ROC curve~\cite{FawcettPRL2006} illustrates the \textit{overall} diagnostic ability of a fault-proneness model.
A ROC curve plots the values of $y\!=\!\TPR$ against the values of $x\!=\!\FPR$ obtained on a dataset by using all possible threshold values $t$ on the fault-proneness model.
A ROC curve is shown in Figure~\ref{fig:ROC}.
The possible set of values of pairs $\langle \FPR, \TPR \rangle$, i.e., the Cartesian product $[0,1] \times [0,1]$, is called the ROC space.

The ROC curve goes through point (1,1) when $t$=0. Setting the threshold $t$=0 implies that all modules are estimated positive, hence \TPR=1 and \FPR=1. When progressively increasing $t$, some module will be estimated negative, thus either decreasing \FPRB (if the module is negative), or decreasing \TPRB (if the module is positive). When $t$=1, all modules are estimated negatives, hence \TPR=0 and \FPR=0. Summarizing, a ROC curve as a function of $t$ always starts at point (1,1) and monotonically decreases until it reaches point (0,0).

\subsection{The ROC Curve of the Random Model}\label{subsec:random}
The random prediction model assigns a random score to each of the $n$ given modules.
When threshold \textit{t} is used to classify modules as positives or negatives, it is expected that $(1\!-\!t)\!\cdot\!n$ models will be estimated positives and $t\!\cdot\!n$ will be estimated negative. On average, of the \APB positives, $(1\!-\!t)\!\cdot\!AP$ are correctly estimated positives, etc.
Table~\ref{tab:RandomModelCM} shows the random model's expected confusion matrix.

\renewcommand{\arraystretch}{1.2}
\begin{table}[htb]
\centering
\caption{Expected confusion matrix for the random model.}
\label{tab:RandomModelCM}
\begin{tabular}{l|c|c|}
 \multicolumn{1}{c}{}     & \multicolumn{1}{c}{Actual Negative}     & \multicolumn{1}{c}{Actual Positive}     \\
\cline{2-3}
Estimated Negative & $\TN_{rnd}=t\!\cdot\!\AN$      & $\FN_{rnd}=t\!\cdot\!\AP$    \\
\cline{2-3}
Estimated Positive & $\FP_{rnd}=(1\!-\!t)\!\cdot\!\AN$   & $\TP_{rnd}=(1\!-\!t)\!\cdot\!\AP$      \\
\cline{2-3}
\end{tabular}
\end{table}
\renewcommand{\arraystretch}{1}
Accordingly, $\TPR_{rnd}=\frac{\TP_{rnd}}{\AP}=\frac{(1\!-\!t)\!\cdot\!\AP}{\AP}=1\!-\!t$, and $\TNR_{rnd}=\frac{\TN_{rnd}}{\AN}=\frac{t\!\cdot\!\AN}{\AN}=t$, hence $\FPR_{rnd}=1\!-\!\TNR_{rnd}=1\!-\!t\!=\!\TPR_{rnd}$.

\begin{figure}[hbt]
\centering
\includegraphics[scale=0.66667]{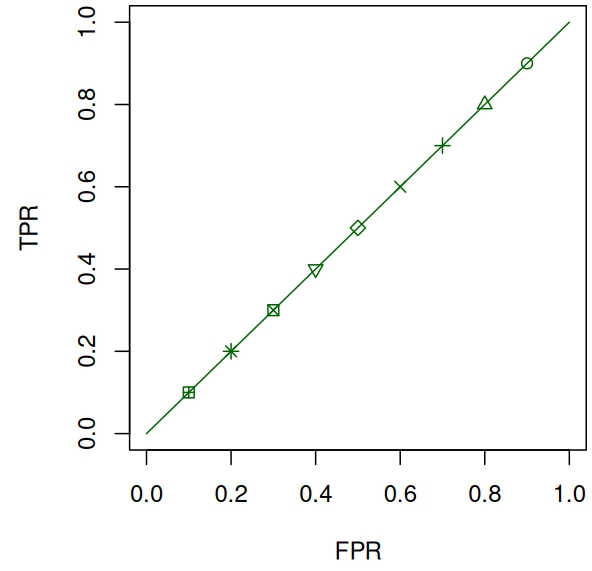}
\caption{The ROC curve of the random model.}
\label{fig:randomROC}
\Description{The ROC curve of the random model is the segment joining points (0,0) and (1,1).}
\end{figure}

Thus, the performance of the random model when the threshold is $t$ is represented in the ROC space by point ($1\!-\!t$,$1\!-\!t$): the ROC curve of the random model is the bisector of the ROC space, i.e., the segment joining points (0,0) and (1,1), as shown in Figure~\ref{fig:randomROC}.

In the figures, points $\circ$, $\triangle$, $+$, $\times$, $\diamondsuit$, $\bigtriangledown$, $\boxtimes$, $*$, $\boxplus$ are associated with thresholds 0.1, 0.2, ..., 0.9, respectively.

The random model provides an easy-to-implement and inexpensive way to make predictions, which does not consider any feature of a module.
This characteristic of the random model qualifies it as a perfect baseline: whatever model is proposed must provide better than random performance. There would be no point in collecting modules' features (including the ground truth) and training a model, if one can get similar or better results by just tossing a coin.

An SDP model with $\TPR\!>\!\TPR_{rnd}$ and $\FPR\!<\!\FPR_{rnd}$ definitely performs better than the random model, being better at classifying both positive and negative modules.
We say that these classifiers perform better than random.
Since $\TPR_{rnd}\!=\!\FPR_{rnd}\!=\!1\!-\!t$, in the ROC space, all points that are located above and to the left of point ($1\!-\!t$, $1\!-\!t$) correspond to classifiers that perform better than random  when threshold $t$ is used to classify modules.

Points that are above and to the right of point ($1\!-\!t$, $1\!-\!t$) represent better than random estimates of positive modules, but worse than random estimates of negative modules. Similarly, points below and to the left of point ($1\!-\!t$, $1\!-\!t$) represent better than random estimates of negative modules, but worse than random estimates of positive modules. That is, these points represent neither better nor worse performance with respect to the random model: they just represent different trade-offs in the performances concerning positive and negative modules.

Clearly, the points that are located below and to the right of point ($1\!-\!t$, $1\!-\!t$) have worse than random performance  when the threshold $t$ is used to classify modules.
These points are below the bisector and represent classifiers that perform worse than the random classifier for both positive and negative modules.

\subsection{The Area Under the ROC Curve (AUC)}\label{subsec:auc}
The \AUCB is the area of the region delimited by the ROC curve and the segments that join point (0,0) with (1,0) and (1,0) with (1,1).
Researchers and practitioners use \AUCB as a performance metric that provides an overall evaluation of the model.
\AUCB is used in Empirical Software Engineering, to evaluate fault-proneness models~\cite{arisholm2007data,beecham2010systematic,catal2012performance,catal2009systematic,singh2010empirical}, as well as for other purposes, and in other domains.

\AUCB is considered representative of the performance of the model because the longer the ROC curve lingers close to point (0,1), which represent the perfect prediction, the better. The closer the ROC gets to point (0,1), the greater the \AUC, in general; therefore, a large value of \AUCB is generally associated with good performance.
\begin{table}[hb]
\centering
\caption{Interpretation of \AUC}
\label{t:EvaluationOfAUC}
\begin{tabular}{rl}
\hline
\AUC range & Evaluation\\
\hline
$\AUC = 0.5$ & totally random, as good as tossing a coin\\
$0.5 < \AUC < 0.7$ & poor, not much better than a coin toss\\
$0.7 \le \AUC < 0.8$ & acceptable\\
$0.8 \le \AUC < 0.9$ & excellent\\
$0.9 \le \AUC$ & outstanding\\
\hline
\end{tabular}
\end{table}

Hosmer at al.~\cite{HosmerLemeshowSturdivant2013} propose the intervals in Table~\ref{t:EvaluationOfAUC} as guidelines to interpret the values of \AUCB as a measure for the discriminating power of a model for all values of threshold $t$.

In addition to being used to evaluate a model, \AUCB is used to compare the performance of two models: the model with the higher \AUCB is preferred, even though it may not be sufficient evidence when the involved ROC curves cross each other~\cite{streiner2007s}.

\section{Fundamental Observations}\label{sec:foundations}

The ROC curve does not show any information concerning thresholds~\cite{ferri2005modifying}. For instance, there may be a part of the curve where many points (corresponding to many threshold values) are concentrated, and other parts where points are more dispersed: the distribution of ROC points with respect to threshold values is not visible, by just looking at the ROC curve. Similarly, given a point of the curve, we cannot in general say what threshold value originated it: the best we can do is to assume that the points close to point (1,1) are associated with low threshold values, and the points close to point (0,0) are associated with high threshold values. With the random model, the situation is quite different, since we know exactly that point ($1\!-\!t$,$1\!-\!t$) is associated with threshold $t$.
 
For a generic threshold $t$, we do not know the relative position of the corresponding point of the model's ROC curve with respect to the random performance, represented by point ($1\!-\!t$, $1\!-\!t$).
The consequence of this observation, along with the considerations given in Section~\ref{subsec:random}, is that $\AUC\!>\!0.5$ and the ROC curve entirely above the bisector are not sufficient conditions to conclude that the considered model is better than random for all threshold values.

As already mentioned, a model is better than random if it classifies both positive and negative instances better than the random model, i.e., if it has both better \TPRB and \FPR.
Since the random model features $\TPR=\FPR=\!1\!-\!t$, a model is better than random if the following condition holds:
\begin{eqnarray}\label{eq:betterThanRandom}
\forall t \in(0,\!1), \TPR(t)\!\ge\!1\!-\!t \wedge \FPR(t)\!\le\!1\!-\!t \wedge \nonumber \\
\exists t \in(0,\!1), \TPR(t)\!>\!1\!-\!t \vee \FPR(t)\!<\!1\!-\!t \   \
\end{eqnarray}
where $\textit{TPR}(t)$ and $\textit{FPR}(t)$ indicate the values of \TPRB and \FPRB obtained with threshold  $t$.

Section~\ref{sec:empirical} shows that condition (\ref{eq:betterThanRandom}) is often not satisfied (i.e., for some $\overline{t}$, it is $\TPR(\overline{t})\!<\!1\!-\!\overline{t}$ or $\FPR(\overline{t})\!>\!1\!-\!\overline{t}$), even if the ROC curve is entirely above the bisector and \AUCB is largely greater than 0.5.

\section{Model Evaluation and Comparison in Practice}\label{sec:empirical}
In this section, we show examples of actual SDP modes---built via state-of-the-art techniques and evaluated against real-life datasets---in which the conditions described in Section~\ref{sec:foundations} do occur.

\subsection{Single Model Evaluation}\label{subsec:oneModel}
\begin{figure}[hbt]
\centering
\includegraphics[scale=5]{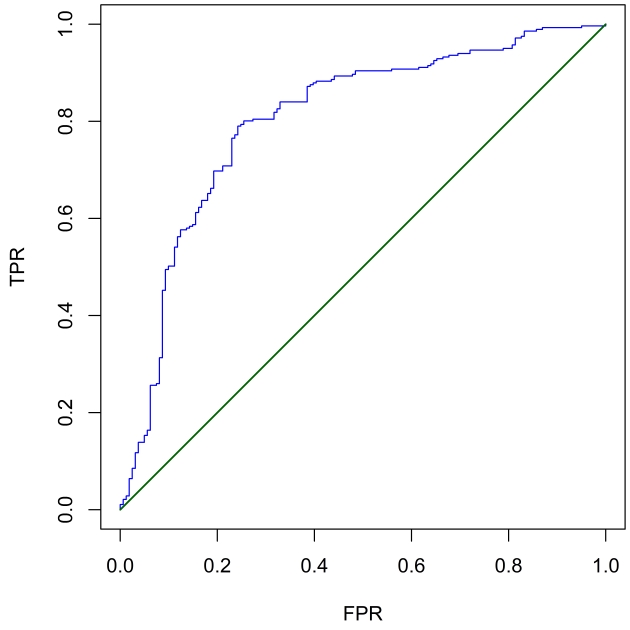}
\caption{The ROC curve of the BLR model for the \texttt{poi 3.0} project (AUC=0.803).}
\label{fig:ROC}
\Description{The ROC curve of the BLR model for the \texttt{poi 3.0} project. The ROC curve features \AUC=0.803.}
\end{figure}

To illustrate the concepts described in Section~\ref{sec:foundations}, we use as an example the leave-one-out cross-validation of Binary Logistic Regression (BLR) defect prediction models obtained from the \texttt{poi 3.0} project dataset~\cite{jureczko2010towards}.
The resulting ROC curve is shown in Figure~\ref{fig:ROC}.

According to the usual criteria, the model's performance is not only largely better than random, since the ROC curve is well above the bisector, but actually good, since \AUC = 0.803.

\begin{figure}[hbt]
\centering
\includegraphics[scale=5]{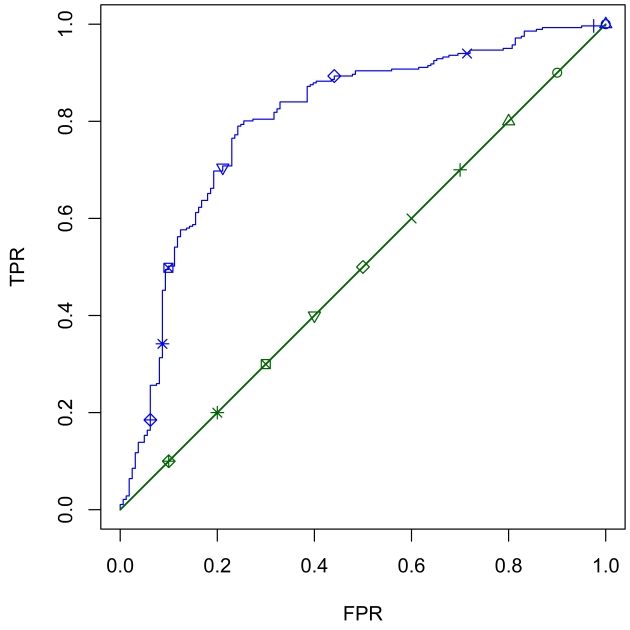}
\caption{The ROC curve of the BLR model for the \texttt{poi 3.0} project. Points representing specific values of the threshold are highlighted.}
\label{fig:ROC_and_p}
\Description{The ROC curve of the RF model for the \texttt{poi 3.0} project, with points corresponding to some threshold values highlighted. The figure shows that for some threshold values the model features FPR worse than random.}
\end{figure}

Figure~\ref{fig:ROC_and_p} represents the same ROC curve shown in Figure~\ref{fig:ROC}, with the points corresponding to thresholds $t$ that are multiple of 0.1 (i.e., 0.1, 0.2, ... , 0.9) highlighted.
Points with the same shape correspond to the same threshold value $t$.
For instance, the points denoted by $\times$ are associated with $t$=0.4 on the ROC curve and on the bisector.

For some threshold values, the model's performances are definitely better than random: for instance, the model's point corresponding to $t$=0.6 (the blue $\triangledown$) is above and to the left of the corresponding random model's point (green point $\triangledown$): when $t$=0.6, the BLR model has both greater \TPRB and smaller \FPRB than the random model.
However, for values of $t\!\le\!0.4$ (e.g., points $\circ$, $\triangle$, $+$ and $\times$) the BLR model has worse \FPRB than the random model: it performs very well in classifying positive (faulty) modules, at the expense of a poor classification of negative (non-faulty) modules.

\begin{figure}[hbt]
\centering
\includegraphics[scale=5]{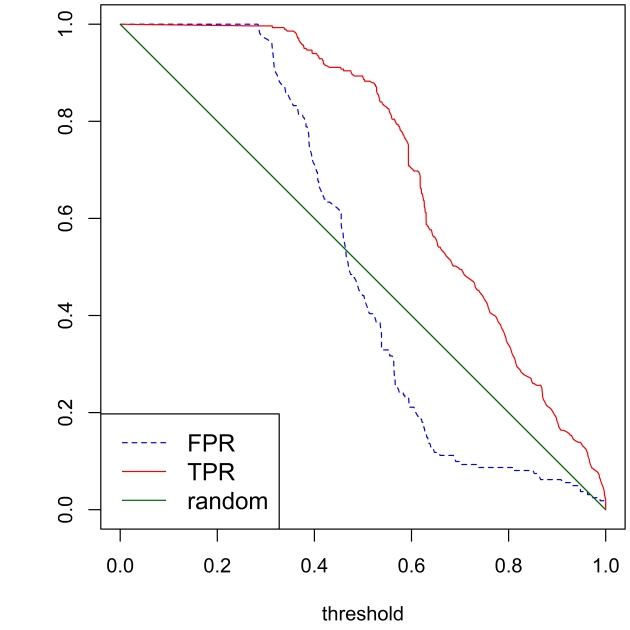}
\caption{TPR and FPR as functions of the threshold, for the BLR model for the \texttt{poi 3.0} project.}
\label{fig:TPR_FPR_p}
\Description{TPR and FPR as functions of the threshold, for the BLR model for the \texttt{poi 3.0} project.}
\end{figure}

To explicitly show the performance associated with a threshold, we can plot \TPRB and \FPRB vs. threshold values.
These plots for our BLR model are in  Figure~\ref{fig:TPR_FPR_p}, where the green line represents both $\TPR_{rnd}$ and $\FPR_{rnd}$.

\FPRB is better than random (i.e., below the $\FPR_{rnd}$ line) only when the threshold is in the (0.464, 0.969) range,
while \TPRB is always better than random (i.e., for all values of the threshold).
Thus, the BLR model features better than random \TPRB \textit{and} \FPRB only when the threshold is in the (0.464, 0.969) range.

\begin{figure}[hbt]
\centering
\includegraphics[scale=5]{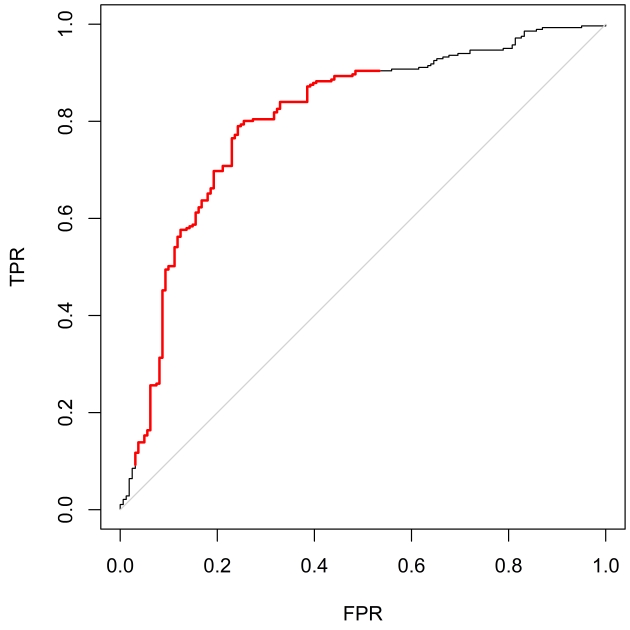}
\caption{The ROC curve for the BLR model for the \texttt{poi 3.0} project: the section of the curve where performance is better than random is highlighted.}
\label{fig:better_section}
\Description{TBD.}
\end{figure}

Figure~\ref{fig:better_section} highlights in red the section of the ROC curve where the performance of the model is better than random with respect to both \TPRB and \FPR. Though this section is only a part of the curve, \AUCB is computed using the entire curve: a practice that can provide misleading indications.

\begin{figure}[hbt]
\centering
\includegraphics[scale=5]{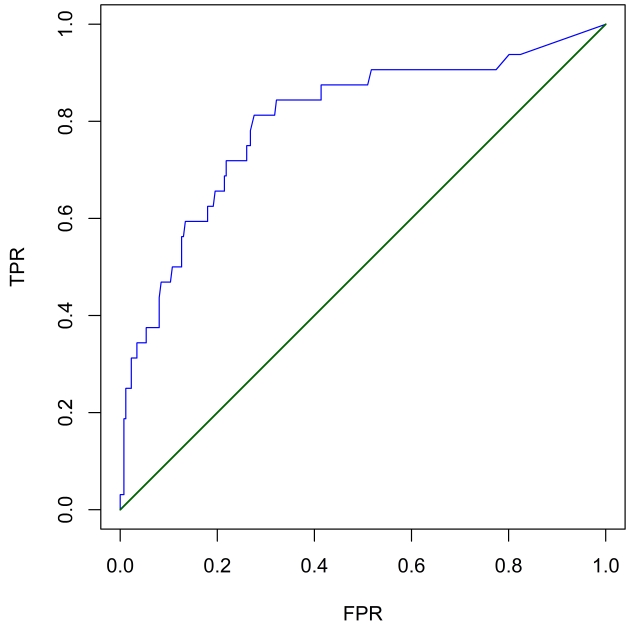}
\caption{The ROC curve of the RF model for the \texttt{ant 1.5} project (\AUC=0.8).}
\label{fig:ROC_ant15_RF}
\Description{The ROC curve of the RF model for the \texttt{ant 1.5} project. The ROC curve features \AUC=0.8.}
\end{figure}

\begin{figure}[hbt]
\centering
\includegraphics[scale=5]{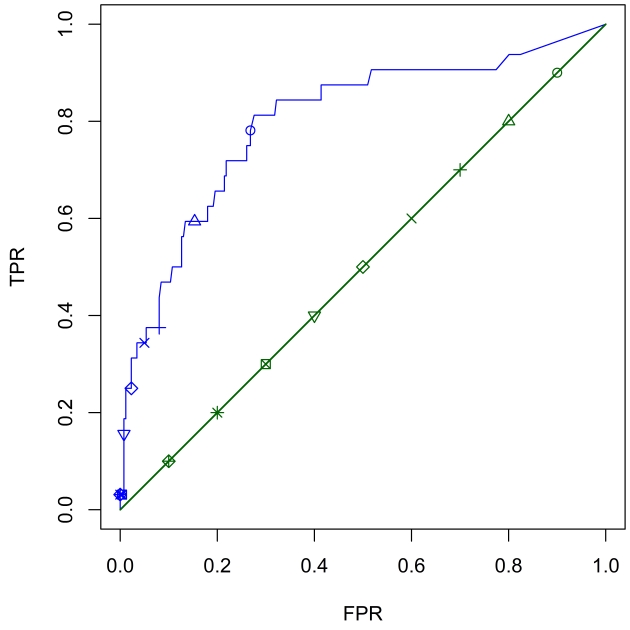}
\caption{The ROC curve of the RF model for the \texttt{ant 1.5} project (AUC=0.8). Points representing specific values of the threshold are highlighted.}
\label{fig:ROC_ant15_RF_p}
\Description{The ROC curve of the RF model for the \texttt{ant 1.5} project. The ROC curve features \AUC=0.8.}
\end{figure}

\begin{figure}[hbt]
\centering
\includegraphics[scale=5]{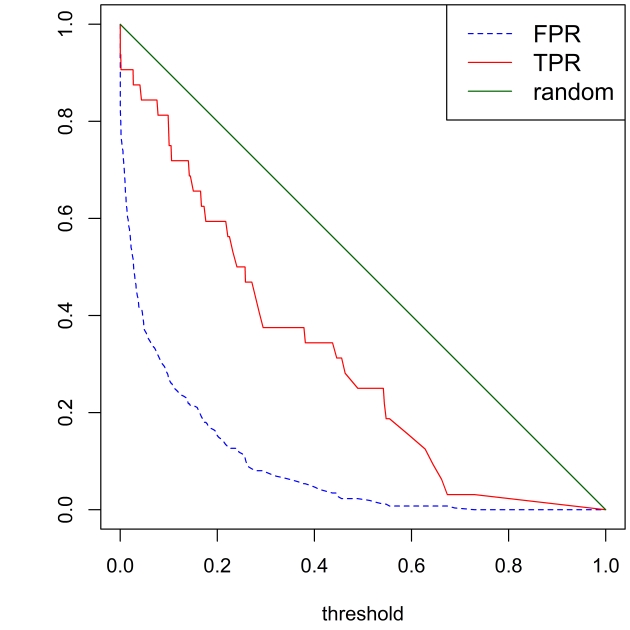}
\caption{TPR and FPR as functions of the threshold, the RF model for the \texttt{ant 1.5} project.}
\label{fig:ROC_ant15_RF_TPR_FPR_vs_t}
\Description{TBD.}
\end{figure}

To illustrate how much \AUCB can be misleading, take the Random Forest (RF) model for project \texttt{ant 1.5}, whose ROC curve (having \AUC=0.8) is shown in Figure~\ref{fig:ROC_ant15_RF}.
Although the value of \AUCB could induce to consider the model as fairly good, by highlighting the points corresponding to specific values of the threshold (as in Figure~\ref{fig:ROC_ant15_RF_p}), it is possible to see that for all the highlighted values of the threshold the model yields worse than random \TPR.
Figure~\ref{fig:ROC_ant15_RF_TPR_FPR_vs_t} shows that for \textit{all} values of the threshold---not just those highlighted in Figure~\ref{fig:ROC_ant15_RF_p}---the model yields worse than random \TPR.
Figure~\ref{fig:ROC_ant15_RF_TPR_FPR_vs_t} also shows that the model has very good \FPR: it is this very good \FPRB that ``pulls'' the ROC curve towards the y axis of the ROC space, causing the ROC curve to be above the bisector and the \AUCB to be high.

Figure~\ref{fig:ROC_ant15_RF_TPR_FPR_vs_t} makes it explicit that the model performs worse (respectively better) then random in classifying positive (respectively negative) modules, for any value of the threshold. This fact cannot be seen by just looking at the ROC curve (Figure~\ref{fig:ROC_ant15_RF}), which is well above the bisector and has a fairly high \AUC.

\subsection{Perfect Models?}\label{subsec:perfect}
A model whose ROC curve goes through point (0,1) has \AUC=1.
Because of the monotonicity property of the ROC curves, this ROC curve is necessarily made of the segment from (1,1) to (0,1) and the segment from (0,1) to (0,0). Such a model is considered perfect~\cite{streiner2007s}.

\begin{figure}[h!]
\centering
\includegraphics[scale=0.6333]{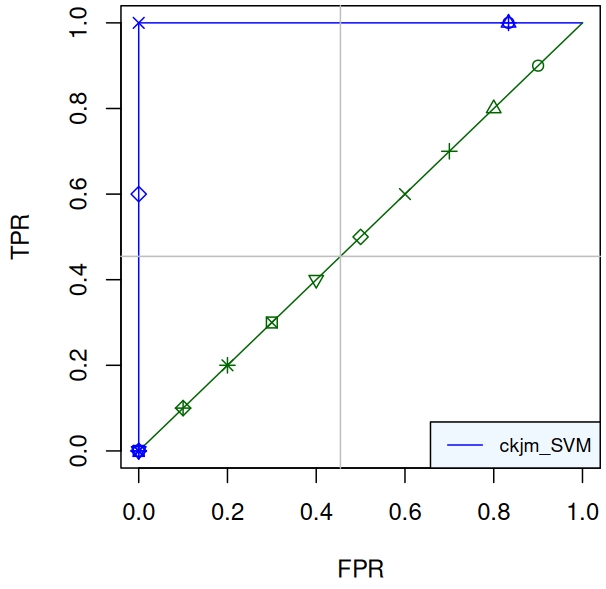}
\caption{ROC curve of SVM model for \texttt{ckjm}.}
\label{fig:ckjm_SVM_ROC_vs_rnd}
\Description{TBD.}
\end{figure}

However, even such a model could provide sub-optimal performance, for some threshold values.
As an example, take the SVM model obtained for project \texttt{ckjm}, whose ROC curve is in Figure~\ref{fig:ckjm_SVM_ROC_vs_rnd}.
The model provides perfect predictions for $t=0.4$ (and for other values of $t$ close to 0.4, not highlighted in Figure~\ref{fig:ckjm_SVM_ROC_vs_rnd}), but several values of $t$ do not yield perfect predictions: e.g., with $t=0.3$, $\FPR>0.8$ (i.e., worse than random), while \TPR=0 for $t\ge0.6$.
This is possible because \AUCB measures only the ranking. Unfortunately, even a  perfect ranking  (i.e., all defective modules preceding non-defective ones) does not exclude that some thresholds yield poor performance.
\AUCB is a property of the ROC curve, which considers \textit{all} the possible thresholds.
So, when \AUC=1, one could be induced to think that the classifier yields perfect predictions with any value of threshold $t$.
This may very well not be the case: \AUC=1 if there is some range $[t_l, t_h]$ such that $t_l\!<\!t\!\le\!t_h$ yields perfect predictions, so that the ROC curve goes through point (0,1).
When $t$ is out of the $[t_l, t_h]$ range, the prediction is not perfect, and possibly not even better than random.
In the case illustrated by Figure~\ref{fig:ckjm_SVM_ROC_vs_rnd}, the perfect prediction range is rather small, being a subset of (0.3, 0.5).

For other models, $t_l$ could be very close to zero and $t_h$ could be very close to one, so that any sensible value of $t$ (i.e., any value not extremely close to zero or one) would yield perfect predictions. Unfortunately, \AUCB does not tell us if we are in this lucky situation or in a situation like the one described above, when we obtain perfect predictions only if we carefully pick a suitable value of $t$.

\subsection{Model Comparison}\label{subsec:comparison}
Model A is considered preferable to model B when model A has greater \AUCB than model B~\cite{FawcettPRL2006}.
However, when two ROC curves cross each other, it is recognized that in some conditions model B could yield better performance than model A, even though $\AUC_A\!>\!\AUC_B$~\cite{FawcettPRL2006,streiner2007s}.
Instead, it is universally accepted that if ROC curve A \textit{dominates} ROC curve B (i.e., ROC curve A is never below ROC curve B and is above it in at least one point), model A is better than model B.
If ROC curve A dominates ROC curve B, then $\AUC_A>\AUC_B$.
However, $\AUC_A>\AUC_B$ does not imply that ROC curve A dominates ROC curve B; in fact, the two curves can cross each-other.

In this section, we show that even when ROC curve A dominates ROC curve B, it is possible that model B is preferable, for some values of the threshold.
As a consequence, it is even more possible that model B is preferable, for some values of the threshold, when  $\AUC_A>\AUC_B$ and curve A does not dominate curve B (i.e., the curve cross each other).

\begin{figure}[h!]
\centering
\includegraphics[scale=5]{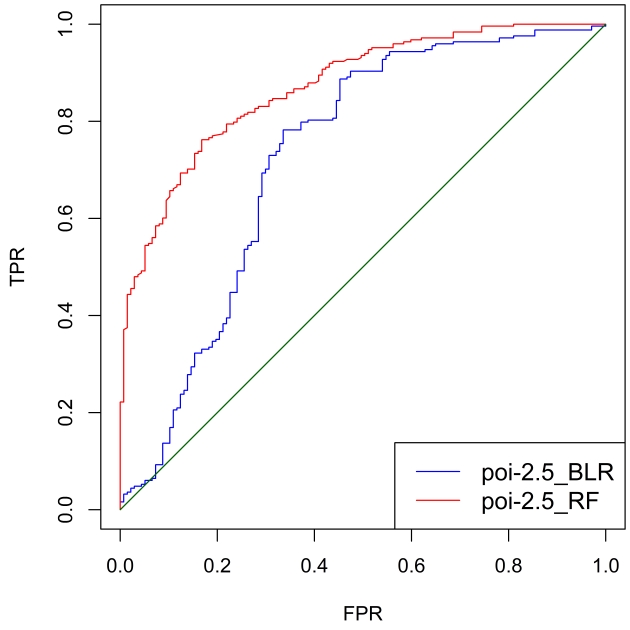}
\caption{ROC curves of RF and BLR models for \texttt{poi 2.5}.}
\label{fig:poi_25_RF_BLR}
\Description{TBD.}
\end{figure}

As an example, let us consider the fault-proneness models for project \texttt{poi 2.5}, obtained via RF and BLR, whose ROC curves are shown in Figure~\ref{fig:poi_25_RF_BLR}.
The ROC curve of the RF model (\AUC=0.87) dominates the ROC curve of the BLR model (\AUC=0.73): based on the traditional interpretation of ROC curves and \AUC, we should conclude that the RF model is better than the BLR model.

However, for each point $P$ of the RF model's curve, there is at least one point $P'$ of the BLR model's curve that achieves either better \TPRB or better \FPRB (but not both).
Noticeably, by just looking at the ROC curves, we do not know whether $P$ and $P'$ are obtained with the same or different fault-proneness threshold values.

Comparing ROC curves without considering the threshold does not seem to provide a really good insight into the relative models' performance.

In analogy to condition (\ref{eq:betterThanRandom}), to state that model A is better than model B,  we can require that
for all thresholds, the  corresponding point of model A has neither \TPRB nor \FPRB worse than the corresponding point of model B, and, for at least one threshold, it has better \TPRB or better \FPR:
\begin{eqnarray}\label{eq:comparison}
\forall t \in(0,\!1), \TPR_A(t)\!\ge\!\TPR_B(t) \wedge \FPR_A(t)\!\le\!\FPR_B(t) \wedge \nonumber \\
\exists t \in(0,\!1), \TPR_A(t)\!>\!\TPR_B(t) \vee \FPR_A(t)\!<\!\FPR_B(t) \   \
\end{eqnarray}

Note that, when condition (\ref{eq:comparison}) holds, ROC curve A dominates ROC curve B, while the reverse is not true in general.

\begin{figure}[hbt]
\centering
\includegraphics[scale=5]{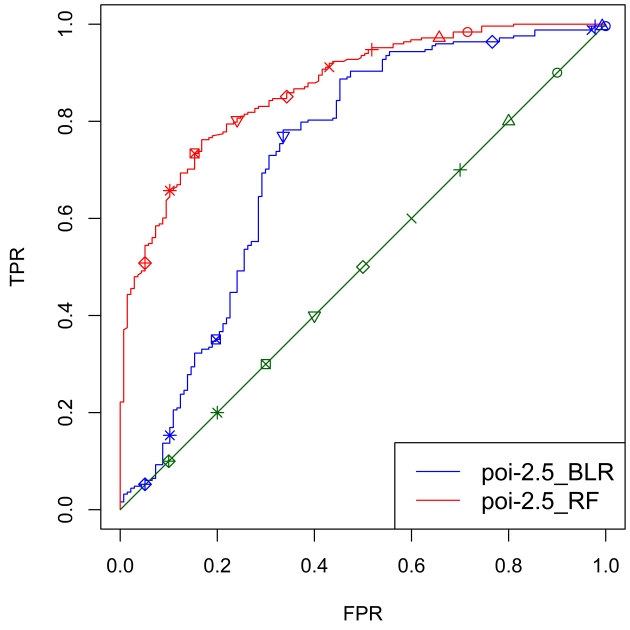}
\caption{Comparison of RF and BLR models for \texttt{poi 2.5}.}
\label{fig:poi_25_RF_BLR_comparison}
\Description{TBD.}
\end{figure}

Many published papers adopted dominance as the criterion used to conclude that the model proposed in the paper outperforms previously published models, both in Software Engineering and in other fields.
However, it may be the case that dominance is satisfied, while condition (\ref{eq:comparison}) is not, as shown in Figure~\ref{fig:poi_25_RF_BLR_comparison}, which represents the same ROC curves given in Figure~\ref{fig:poi_25_RF_BLR}, with the points corresponding to specific threshold values highlighted.

When $t$=0.6 (point $\triangledown$) the RF model is better than the BLR model with respect to both \TPRB and \FPR.
However, for multiple threshold values, it is not so: when $t$=0.5 (point $\diamondsuit$) the BLR model achieves better \TPRB and worse \FPR, while with $t$=0.8 (point $\ast$) the BLR model achieves slightly better \FPRB and worse \TPR.
For several threshold values, the two models provide different trade-offs concerning the classification of positive and negative modules, neither of them being better at classifying both positives and negatives.

\begin{figure}[hbt]
\centering
\includegraphics[scale=5]{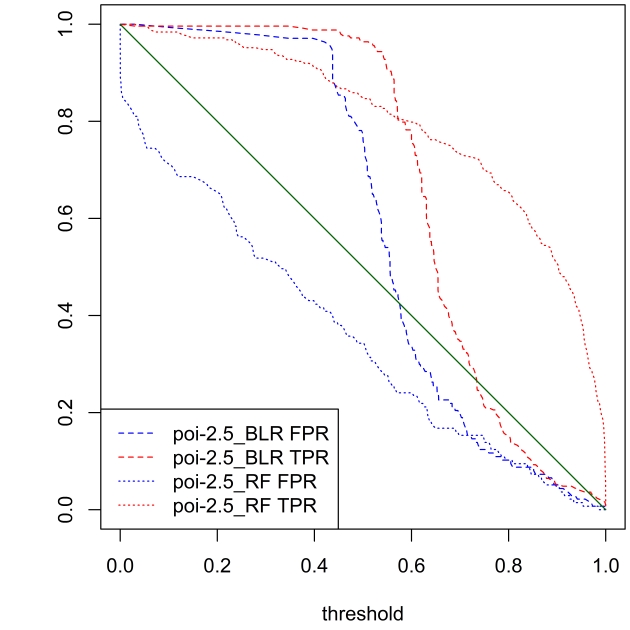}
\caption{Comparison of RF and BLR models for \texttt{poi 2.5}.}
\label{fig:comp_poi25_BLR_RF_TPR-FPR_t}
\Description{TBD.}
\end{figure}

It is interesting to note that, although the ROC curves do not cross each other, both the $\TPR(t)$ and $\FPR(t)$ curves cross each other (as shown in Figure~\ref{fig:comp_poi25_BLR_RF_TPR-FPR_t}), thus clearly indicating that neither model is better than the other for all threshold values.

\begin{figure}[hbt]
\centering
\includegraphics[scale=5]{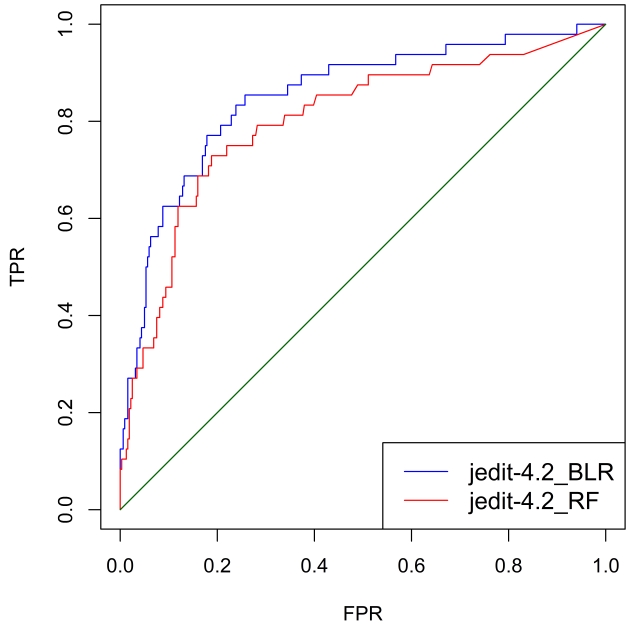}
\caption{ROC curves of RF and BLR models for \texttt{jedit 4.2}.}
\label{fig:jedit42_RF_BLR}
\Description{TBD.}
\end{figure}

\begin{figure}[hbt]
\centering
\includegraphics[scale=5]{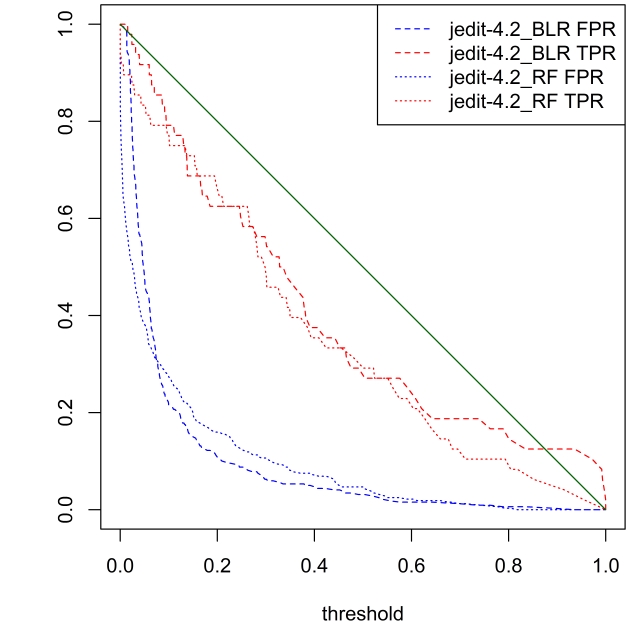}
\caption{\TPRB and \FPRB of RF and BLR models for \texttt{jedit 4.2} as functions of the threshold.}
\label{fig:comp_jedit42_BLR_RF_TPR-FPR_t}
\Description{TBD.}
\end{figure}

Let us now consider the RF and BLR models for \texttt{jedit 4.2}.
Figure~\ref{fig:jedit42_RF_BLR} shows that the ROC curve of the BLR model (\AUC=0.85) dominates the ROC curve of RF model (\AUC=0.8).
However, Figure~\ref{fig:comp_jedit42_BLR_RF_TPR-FPR_t} clearly shows that both models yield \TPRB values that are generally largely worse than random, hence probably neither model would be considered usable by a practitioner.

\section{The Effect of Imbalance}\label{sec:imbalance}
In imbalanced datasets, probabilities are biased towards the most common event~\cite{cramer1999predictive,HosmerLemeshow2000}.
Thus, a large prevalence of positive modules ($\rho$ close to 1) leads to maximizing \TPR, at the expense of having poor \FPR, for most values of the threshold.
Similarly,  a very small prevalence ($\rho$ close to 0) leads to maximizing \FPR, at the expense of having poor \TPR, for most values of the threshold.
In these conditions, ROC curves tend to remain well above the bisector, thus showing fair or even good performance, characterized by decent values of \AUC~\cite{lavazza2023reliability}.

\begin{figure*}[hbt]
\centering
\begin{tabular}{cc}
\includegraphics[scale=5]{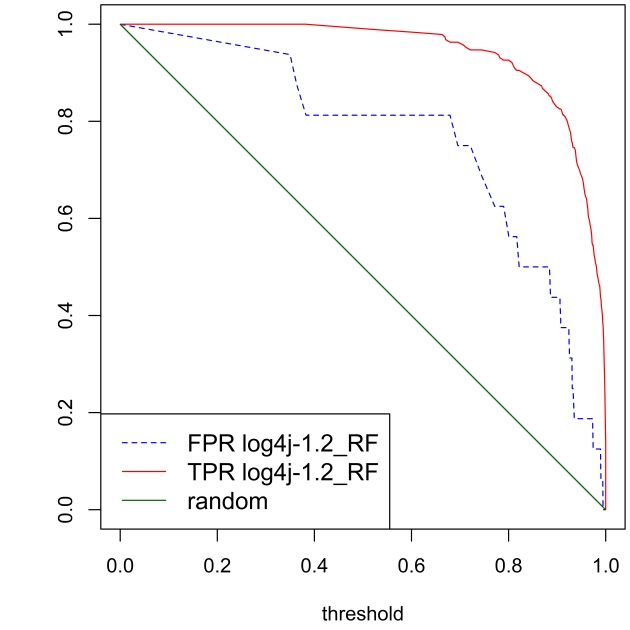} &
\includegraphics[scale=5]{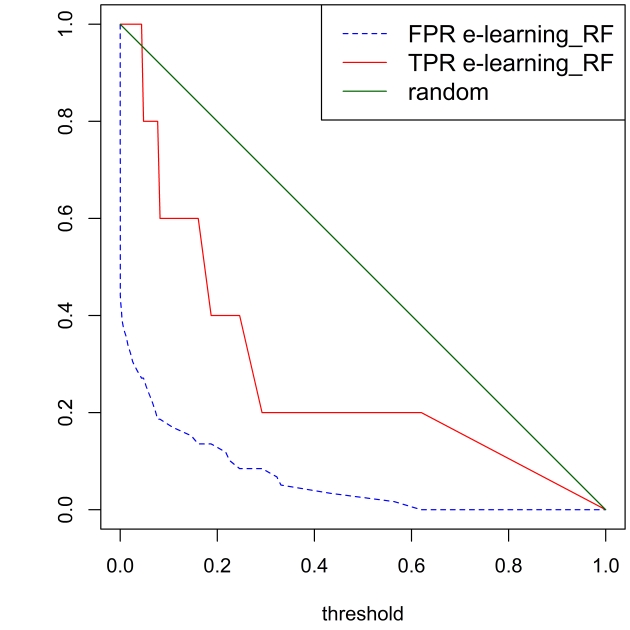} \\
\end{tabular}
 \caption{\TPR($t$) and \FPR($t$), when prevalence $\rho$ is high (\texttt{log4j 1.2}, left) and low (\texttt{e-learning}, right).}
\label{fig:low_high_rho}
\Description{xxxx}
\end{figure*}

Let us consider the RF models obtained for projects \texttt{log4j 1.2} ($\rho=0.92$) and \texttt{e-learning} ($\rho=0.08$).
Figure~\ref{fig:low_high_rho} shows the \TPRB and \FPRB obtained by the mentioned models, as functions of the threshold.
It is apparent that the two models achieve quite different performances: \texttt{log4j 1.2} has very good \TPRB for threshold values up to around 0.8, while \FPRB is always largely worse than random.
Instead, \texttt{e-learning} has very good \FPRB for threshold values above 0.1, while \TPRB is often largely worse than random.

The situation shown in Figure~\ref{fig:low_high_rho} is hardly surprising. When the positive modules are very prevalent, it is easy to identify those modules; the difficult part is identifying them without misclassifying too many negative modules.
The same type of reasoning applies for the prevalence of negative elements.

\begin{figure*}[hbt]
\centering
\begin{tabular}{cc}
\includegraphics[scale=5]{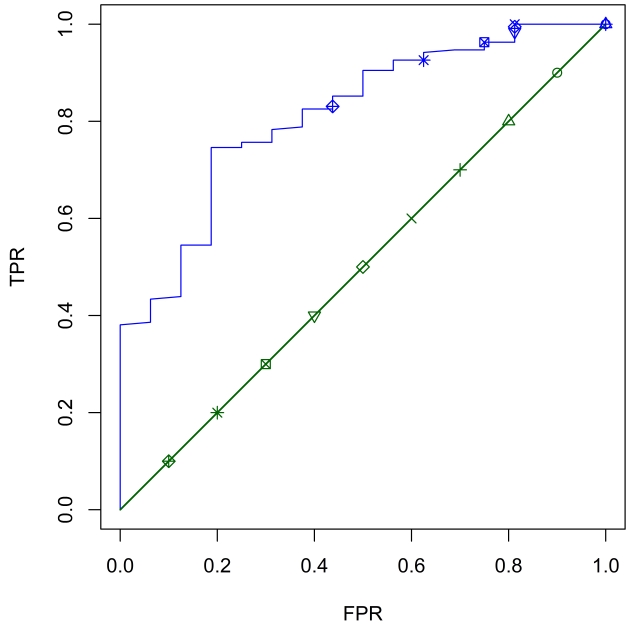} &
\includegraphics[scale=5]{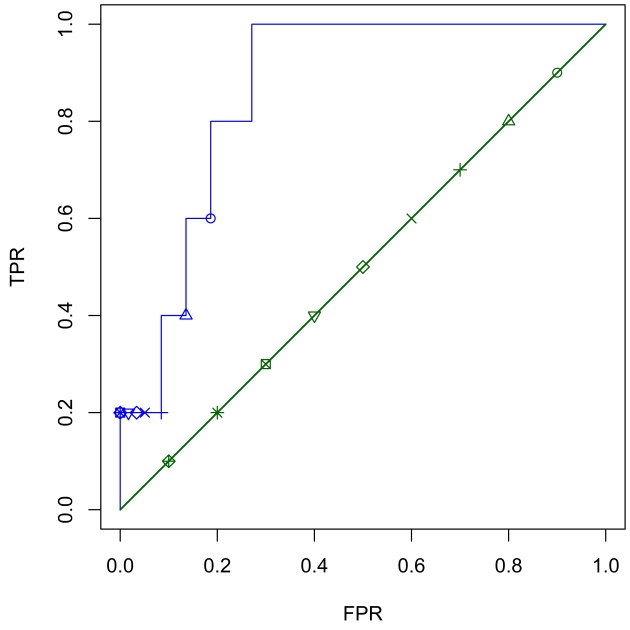} \\
\end{tabular}
 \caption{ROC curves with thresholds highlighted, when prevalence $\rho$ is high (\texttt{log4j 1.2}, left) and low (\texttt{e-learning}, right).}
\label{fig:low_high_rho_ROC}
\Description{xxxx}
\end{figure*}

Figure~\ref{fig:low_high_rho} shows that the considered models do not perform well: both of them achieve very good performance with the prevalent class of modules, but misclassify the minority class modules.
Nonetheless, the ROC curves of both models are reasonably good, as shown in Figure~\ref{fig:low_high_rho_ROC}: the corresponding \AUCB values are 0.81 for \texttt{log4j 1.2} and 0.86 for \texttt{e-learning}.

Figure~\ref{fig:low_high_rho_ROC} shows the effects of imbalance, which are not visible in regular ROC curves.
The largest part of the ROC curve of the model for \texttt{log4j 1.2} (which has prevalence $\rho=0.917$) is made of points obtained with threshold $t\!>\!0.9$.
Instead, the largest part of the ROC curve of the model for \texttt{e-learning} (which has prevalence $\rho=0.078$) is made of points obtained with threshold $t\!<\!0.1$.
This phenomenon is consistent with the observation given above, that when positive modules are largely prevalent most modules are estimated positive, and vice-versa.

In conclusion, in case of severely imbalanced datasets, \AUCB appears definitely overoptimist, while ROC curves hide worse than random performance concerning either \TPRB or \FPR.

\section{The Empirical Study}\label{sec:study}

One could wonder how often \AUCB provides unreliable indications: to clarify this issue, we carried out an empirical study.
For this study, we employed Binary Logistic Regression (BLR) and Random Forest (RF) models, as these techniques compute a score for each module in the [0,1] range, which is interpreted as a probability of a module to be positive.

We trained models on 65 datasets from the Jureczko\&Madeyski (J\&M) collection~\cite{jureczko2010towards} and 11 datasets from the NASA collection~\cite{shepperd2013data}. The descriptive statistics of the datasets are given in Table~\ref{tab:datasets}.

\begin{table}[hbt]
	\caption{Descriptive statistics of the datasets used in the empirical study.}
	\label{tab:datasets}
	\begin{tabular}{lllllll}
		\hline
		& \multicolumn{3}{c}{\# modules} & \multicolumn{3}{c}{\% faulty modules} \\
		& Mean         & Min        & Max         & Mean           & Min           & Max           \\ \hline
		NASA & 857          & 124        & 1952        & 15.2\%         & 1.8\%         & 35.5\%        \\
		J\&M & 283.1        & 10         & 965         & 34.1\%         & 2.2\%         & 98.8\%        \\ \hline
	\end{tabular}
\end{table}

Of the many metrics provided in both the J\&M and NASA datasets, only a subset appear relevant to estimate module defectiveness: we used only those metrics (8 from the J\&M datasets and 9 from the NASA datasets) to build SDP models.
We built models using groups of up to three metrics as independent variables~\cite{fu2017easy}.
Therefore, we obtained $\binom{8}{3}\!=\!56$ models for each dataset from the J\&M collection, and $\binom{9}{3}\!=\!84$ models for each dataset from the NASA collection.
The estimates were evaluated via Leave-One-Out Cross Validation.

All the computations were carried out using the \texttt{R} environment. Specifically, the RF models were built using the \texttt{RandomForest} library. This library allows for the tuning of the \texttt{mtry} hyperparameter, but this was not needed since we considered no more than three features at a time. We set the number of trees to 1,500, since using more trees did not seem to improve performance.

To evaluate to what extent the problems described in Sections~\ref{sec:foundations} and~\ref{sec:empirical} affect SDP models, we computed the following numbers:
\begin{enumerate}
	\item
	How many models were obtained.
	\item
	How many models feature a ROC curve completely above the bisector (hence, have $\AUC\!>$0.5).
	\item
	How many models have $\AUC\ge0.8$.
\end{enumerate}

Of the models at points (2) and (3) above, we checked how often some points of the curve have worse than random \TPRB or \FPR.

Concerning model comparison, we selected the cases where there is a clear winner, according to the traditional interpretation of the ROC curves, i.e., a model's ROC dominates the other models' ROC curves. Then, we checked if the winner model is actually better than the other models with respect to both \TPRB and \FPR, for all threshold values.

\subsection{Results}

As mentioned before, we built 56 models for each dataset from the J\&M collection, and 84 from the NASA collection. We used two Machine Learning techniques, so we obtained a total of $2\times((56\times65) + (84\times11))=$ 9,128 models.
Of these models, 8,651 have $\AUC\!>\!0.5$;
4,934 models generate ROC curves that have no points below the bisector;
2,304 models are associated with ROC curves that are strictly above the bisector (except for points (0,0) and (1,1)).

Of the 2,304 models whose ROC curves are completely above the bisector, only 114 (around 5\%)  have performance that is better than random with respect to both \TPRB and \FPRB for all thresholds.

1,504 of the 2,304 models with ROC curves completely above the bisector have $\AUC\!\ge\!0.8$, but only 100 of them (6.6\%) are better than random with respect to both \FPRB and \TPRB for all thresholds.

We also compared all the pairs of models obtained using the same dataset.
For each model and dataset, we generated 56 models for  J\&M datasets and 84 for NASA datasets using two modeling techniques, thus obtaining 112 and 168 total models per dataset for the two collections. Hence, we performed $=\frac{112\times 111}{2}$=6,216 pair-wise comparisons for J\&M datasets and $=\frac{168\times 167}{2}$=14,028 comparisons for NASA datasets, for a grand total of $(6,216\times 65) + (11\times 14,028)$=558,348 pairwise comparisons.

In 82,489 cases, one ROC curve dominates the other. However, only in 901 cases (1.1\%) one curve has better \TPRB and \FPRB for all threshold values.
If we only consider the comparisons in which one curve dominates the other, and the difference in \AUCB values is greater than 0.2, we still have that only in 823 comparisons out of 29,845 (2.8\%) one curve has better \TPRB and \FPRB than the other for all thresholds.

In conclusion, both when evaluating a single SDP model and when comparing SDP models,
the traditional interpretations of the ROC curve and the \AUCB are very frequently misleading.

\subsection{Threats to Validity}\label{sec:threats}

\textit{External validity.}
For the empirical study, we employed two widely used dataset collections, and widely used modeling techniques (BLR and RF), whose predictions can be interpreted as probabilities.
Other models, datasets, and contexts (such as just-in-time SDP) could have been considered.
However, we argue that similar observations could have been made with other probabilistic classifiers and datasets, thus limiting threats to external validity.

\textit{Internal validity.}
We trained and tested our models via mature and stable libraries of the \texttt{R} environment (\texttt{pROC} for ROC curves, \texttt{stats} for BLR models, and \texttt{RandomForest} for RF models).

\textit{Construct validity.}
We study the construct validity of \AUCB as a sensible performance metric for SDP models. Thus, we detect and discuss its construct validity issues throughout the article. As we stated in Section \ref{subsec:roc}, we consider SDP models that estimate the probability that a software module is faulty. However, several Machine Learning algorithms return a score that may not be reliably taken as a probability. If so, it is recommended that the output of these models be calibrated to represent probabilities, for instance with Platt scaling~\cite{platt1999probabilistic}.

\textit{Conclusion validity.}
We do not claim that a model with performance worse than random for some threshold values should be always discarded. We point out, however, that a good ROC curve with a good \AUCB does not necessarily represent a model that is good (or even perfect, as in Figure~\ref{fig:ckjm_SVM_ROC_vs_rnd}) in general.
Similarly, a model with a higher \AUCB than another, or even with a ROC curve that dominates another model's, is not necessarily better.
A model should be selected while keeping the preferred threshold range in mind.

\section{Guidelines}\label{sec:usage}
Here, we provide some suggestions about the evaluation of SDP models in a ``threshold-aware'' way.
We only consider the comparison of models: this is not a limitation, since the evaluation of a single  SDP model is in fact a comparison with the random model.

Let us consider SDP models A and B; let $ROC_A$ and $ROC_B$ be the ROC curves of the two models, and let ($\FPR_A(t), \TPR_A(t)$), respectively ($\FPR_B(t), \TPR_B(t)$), be the point of $ROC_A$, respectively, $ROC_B$, when the threshold's value is $t$.

\begin{figure}[hbt]
\centering
\includegraphics[scale=0.35]{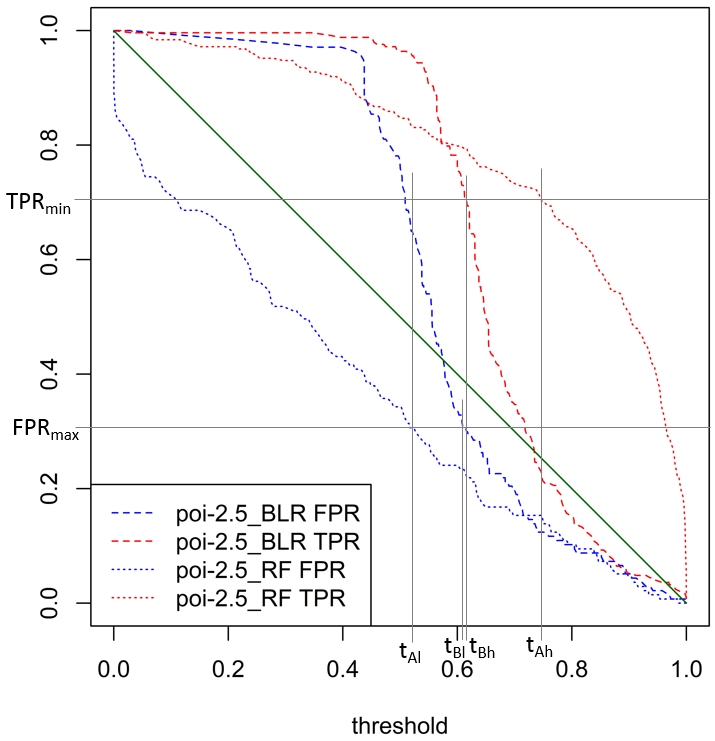}
\caption{Comparison when condition (\ref{eq:comparison}) does not hold.}
\label{fig:selection_criteria}
\Description{xxxxx}
\end{figure}

If condition (\ref{eq:comparison}) holds, there is no doubt that A is preferable.
Otherwise, it is necessary to consider which model provides the best trade-off between \TPRB and \FPR,
for what threshold values.
There are many ways to make such evaluation. Here we illustrate a possible procedure, with the help of Figure~\ref{fig:selection_criteria}, where model A and B are the RF and BLR models, respectively, for project \texttt{poi 2.5}.

Based on context-specific considerations, one can define the minimum acceptable
performance in terms of $\TPR_{min}$ and $\FPR_{max}$: in Figure~\ref{fig:selection_criteria}, $\TPR_{min}\!=\!0.7$ and $\FPR_{max}\!=\!0.3$.
In the threshold range [$t_{Al}$, $t_{Ah}$] model A has $\TPR_A\!\ge\!\TPR_{min}$ and $\FPR_A\!\le\!\FPR_{max}$. Similarly, model B has acceptable performance in [$t_{Bl}$, $t_{Bh}$].
It is now possible to identify the ranges where both models are better than $\TPR_{min}$ and $\FPR_{max}$ (if any) and compare the performance of A and B in those ranges.

The latter comparison can be carried out according to multiple criteria.
A first criterion is that a model that has acceptable performance in a wide range (like that of the RF model A in Figure~\ref{fig:selection_criteria}) is preferable to a model that performs acceptably in a narrow range, because the narrower the range, the more likely the chosen threshold will be outside the range, thus yielding poor predictions.
Another criterion consists in evaluating the relative performance in the acceptable ranges: e.g., Figure~\ref{fig:selection_criteria} shows that model A (RF) has both better \TPRB and better \FPRB in the entire intersection of [$t_{Al}$, $t_{Ah}$] and [$t_{Bl}$, $t_{Bh}$] (though not in the entire [$t_{Al}$, $t_{Ah}$] range).

An alternative procedure is applicable when the cost of false negatives (e.g., the cost due to releasing a defective module) and false positives (wasted effort) is known. In such case, instead of looking at $\TPR(t)$ and $\FPR(t)$, as in Figure~\ref{fig:selection_criteria}, we can plot the cost as a function of $t$: a model that minimizes the cost, especially for a large range of threshold values, is definitely preferable.

\section{Related Work}\label{sec:related}
Machine Learning research has criticized ROC curves because they only show how well samples have been ranked, and ignore the distribution of probabilities among samples~\cite{ferri2005modifying}.

In their audit about Machine Learning experimentation in Software Engineering, Destefanis et al.~\cite{destefanis2026audit} show that \AUCB is the second most popular performance metric. They note that \AUCB is ``chance-anchored''~\cite{chicco2021matthews}, since the value of \AUCB of a binary classifier can immediately be compared to the value of the random classifier, which is 0.5. In that, \AUCB is akin to an effect measure, since its value tells ``how far'' the classifier is from being totally random~\cite{GiGiSandroFMvsPhi}.

In a medical paper, Calster et al. point out how ROC curves and \AUCB can be difficult to interpret when trying to measure models' improvement when adding new markers~\cite{van2014sensitivity}. They make similar observations to those in our paper, showing how, even when the ``improved'' ROC curve is completely above the base curve, there are thresholds in which the performance of the new curve are actually lower. The paper criticizes other metrics specifically used to measure improvement, indicating how making more ``threshold-aware'' evaluations can be useful in different applications.

A recent paper in the Journal of Epidemiology reiterates the issue, arguing that researchers should avoid showing ROC curves without any information on threshold values~\cite{verbakel2020roc}. They also suggest the use of ``classification plots,'' i.e., plots of \TPRB and \FPRB values over threshold values, like the one depicted in Figure~\ref{fig:ROC_ant15_RF_TPR_FPR_vs_t}. However, some considerations seem questionable, at least for fault-proneness models. Specifically, they argue that ROC curves should not be used because curves with equal \AUCB (obtained on the same test dataset) can vary a lot, adding ``complexity without adding useful information.'' We think that, even without explicit information on predicted probabilities, the shape of a ROC curve can be informative, especially when comparing curves with the same \AUC.

It has been shown that the values of any performance metric (e.g., the Matthews Correlation Coefficient, the F-score, etc.) can be plotted in the ROC space~\cite{MorascaLavazzaEMSE2020,lavazza2023reliability,lavazza2025software}: it is thus possible to evaluate the position of every point of a ROC curve both with respect to the corresponding random performance and with any traditional performance metric.
It is also possible to plot the value of any performance metric as a function of the threshold, and compare it with the corresponding random model's plot.

\section{Conclusions}\label{sec:conclusions}
We showed that ROC curves and their associated \AUCB can yield misleading indications concerning the performance of binary predictive models.
In fact, although ROC curves that are entirely above the bisector of the ROC space are traditionally considered better than the random classifier, it is often the case that for some threshold values either \TPRB or \FPRB are worse than the random model's.

Similarly, when comparing models, a model whose ROC curve is completely above the other model's curve (hence, has higher \AUC) may have worse \TPRB or \FPR, for some thresholds.

Many research papers in the Software Engineering area (including the majority of papers dealing with SDP models~\cite{moussa2022use}) use the \AUCB to evaluate models.
However, as we have shown, the \AUCB is not a reliable indicator of the performance of classifiers.
As a consequence, it is possible that many evaluations reported in research papers are not correct.
Reconsidering the usage of \AUCB  appears urgent.
To this end, we suggest using more reliable indicators, like the graphs that show how a performance metric depends on the threshold value.
These graphs can be used to show together $\TPR(t)$ and $\FPR(t)$ (as in Figure~\ref{fig:selection_criteria}), to appreciate the trade-off achieved in correctly classifying positive and negative modules, for any possible (and reasonable) value of $t$.
In case the relative cost of false positives and false negatives is known, it is also possible to plot the cost function and choose the model accordingly.

\begin{acks}
This work has been partly supported by the ``Fondo di Ricerca d'Ateneo'' funded by the Universit\`a degli Studi dell'Insubria.
\end{acks}

\subsection*{Online Resources}
A package with all the ROC curves we generated and the data we collected
is available at \url{https://doi.org/10.6084/m9.figshare.31100191}.

\bibliographystyle{ACM-Reference-Format}
\bibliography{theBib}

\end{document}